\documentclass[aps,reprint,footinbib]{revtex4-1}
\usepackage{dcolumn,bm,amsmath,amssymb,amsfonts,graphicx,natbib}
\usepackage[utf8]{inputenc}
\usepackage[T1]{fontenc}
\usepackage{mathptmx}
\usepackage{caption,subcaption}
\usepackage[group-separator={,}]{siunitx}
\newcommand{\be}{\begin{equation}}
\newcommand{\ee}{\end{equation}}

\begin{document}
\title{Turbulence model reduction by deep learning}
\author{R. A. Heinonen}
\author{P. H. Diamond}
\affiliation{University of California San Diego, La Jolla, California 92093}
\date{\today}
\begin{abstract}
A central problem of turbulence theory is to produce a predictive model for turbulent fluxes. These have profound implications for virtually all aspects of the turbulence dynamics. In magnetic confinement devices, drift-wave turbulence produces anomalous fluxes via cross-correlations between fluctuations. In this work, we introduce a new, data-driven method for parameterizing these fluxes. The method uses deep supervised learning to infer a reduced mean-field model from a set of numerical simulations. We apply the method to a simple drift-wave turbulence system and find a significant new effect which couples the particle flux to the local \emph{gradient} of vorticity. Notably, here, this effect is much stronger than the oft-invoked shear suppression effect. We also recover the result via a simple calculation. The vorticity gradient effect tends to modulate the density profile. In addition, our method recovers a model for spontaneous zonal flow generation by negative viscosity, stabilized by nonlinear and hyperviscous terms. We highlight the important role of symmetry to implementation of the new method.
\end{abstract}

\maketitle

Interest in turbulence --- the principal ``unsolved'' problem in classical physics --- is driven both by the challenge of understanding the strongly nonlinear dynamics, and by the need for tractable models of turbulent transport. The study of such model reduction began with Prandtl's mixing-length theory approach to pipe flow transport and profile formation \cite{prandtl}, which has been extended to thermal transport and boundary layers \cite{landau}, heat transfer \cite{pope}, stellar structure \cite{padmanabhan}, ocean surface layer mixing \cite{phillips}, and accretion disk dynamics \cite{shakura}. 

One problem that appears frequently is that of accurately modeling transport in different channels, such as heat and momentum, particles and heat, etc. It is intuitively appealing to consider such turbulent transport by a matrix flux-gradient relation 
\be
\label{eq:fluxgrad}
\Gamma_{\alpha} = -D_{\alpha\beta} \nabla \xi_\beta
\ee 
where $\Gamma_\alpha$ is the vector of turbulent fluxes, $\nabla \xi_\alpha$ is the vector of driving gradients or thermodynamic forces, $D_{\alpha \beta}$ is the matrix of transport coefficients, and the indices refer to transport channels. In many cases, some elements of $\mathbf{D}$ can be negative, as relaxation in some channels can drive up-gradient fluxes in others. One such system is drift-wave turbulence in magnetically confined plasmas \cite{dupree67}, where up-gradient transport processes (i.e.\ zonal flow formation) and inward density pinch \cite{kadomtsev66} (akin to chemotaxis) are familiar. In some systems, the challenge of calculating $\mathbf{D}$ is principally one of \emph{simultaneous} determination of the cross-phases between the various channels $\xi_\alpha$. To date, theory has not been especially successful in confronting the problem of predicting cross-phases in multiple channels. For example, most models claiming to calculate $\mathbf{D}$ are based on quasilinear theory \cite{vvs61,vvs62}, the use of which is frequently beyond justification.

In this work, we introduce a new, data-driven method based on deep supervised learning \cite{hinton_review} which infers a mean-field model for the cross-phases from direct numerical simulation (DNS). The mean-field model self-consistently describes the coupled radial dynamics of the principal mean fields (i.e.\ profiles) of physical interest. This method, a form of nonlinear, nonparametric regression, does not rely on any approximations besides the applicability of local mean field theory, and it can be used to either validate existing models or probe for new physics.

As a test of concept, we use direct numerical simulation of the 2-D Hasegawa-Wakatani (HW) system \cite{hw83,hw84,numata} --- a variant of the quasigeostrophic or Charney-Hasegawa-Mima system \cite{charney,hm77} --- to train a deep neural network (DNN) which outputs the local turbulent particle flux and poloidal momentum flux (Reynolds stress) as a function of local mean gradients, flow properties, and turbulence intensity. Exact symmetries are exploited to select independent variables and constrain the model. 

The key results of this paper are as follows. The DNN infers a model for the turbulent particle flux of the form $\langle \tilde v_x \tilde n \rangle \simeq \varepsilon(-D_n\partial_x N+D_u \partial_x U)$ where $\varepsilon$ is the turbulence intensity, $N$ and $U$ are respectively the mean density and vorticity, and $D_n$ and $D_u$ are constants. [Throughout this paper, $\langle \cdot \rangle$ will represent an average over (poloidal and toroidal) directions of symmetry and a tilde will represent the local deviation from this average.] This form is valid when higher-order effects are negligible. The first term is the familiar turbulent diffusion which tends to relax the driving gradient. The second, proportional to the gradient of vorticity, is non-diffusive and previously unreported. The vorticity gradient effect tends to modulate the profile in the presence of a quasiperiodic zonal flow. In contrast, there is relatively weak direct dependence of the flux on the vorticity (shear) itself, contradicting the conventional wisdom that turbulent transport is directly suppressed by the shear.

Meanwhile, the DNN uncovers a Reynolds stress closure for the generation of zonal flow. Explicitly, we find at leading order $\langle \tilde v_x \tilde v_y\rangle\simeq\varepsilon (-\chi_1 U+\chi_3 U^3-\chi_4 \partial_x^2 U)$, which agrees with previous theoretical work \cite{diamond2005} identifying negative viscosity \cite{starr} as the mechanism for spontaneous zonal flow generation. The resulting mean vorticity evolution equation resembles the Cahn-Hilliard equation \cite{cahn-hilliard}, a model of phase separation. We also find higher-order corrections, including dependence on the gradients of $N$ and $U$. These results form the basis of a novel, reduced 1-D model for the turbulent dynamics. 

A generalization of Eq.~(\ref{eq:fluxgrad}) expresses the local (radial) flux as an arbitrary function of local mean variables such as temperature or density gradients, the $E\times B$ flow shear, and some measure of the turbulence intensity. Formally, to construct such a local mean-field model, one choses a collection of $n$ spatiotemporally-varying fields $\psi_i(\mathbf{x},t)$ and seeks a map
\begin{equation}
\label{eq:map}
\mathcal{M}_\xi : (\langle \psi_1 \rangle,\dots, \langle \psi_n \rangle)|_{r_0,t_0} \mapsto \langle  \tilde v_r (r_0,t_0) \tilde \xi(r_0,t_0) \rangle
\end{equation}
outputting the turbulent flux of $\xi$ at a radius and time $(r_0,t_0)$.

Our method selects the local mean-field model that best explains the dynamics, according to a loss function which quantifies the prediction error. It leverages deep learning's resilience to the large amounts of noise inherent to turbulence \cite{rolnick}, as well as its ability to model arbitrary nonlinear, multivariate functions \cite{hornik91,leshno93,lu2017}.

 We apply the method to a particularly simple description of resistive drift-wave turbulence, the (modified) 2-D Hasegawa-Wakatani (HW) model in a periodic slab \cite{hw83,hw84,numata}:
\begin{align}
\label{eq:hw}
\partial_t n + \{ \phi, n \}&= \alpha (\tilde \phi - \tilde n) - D \nabla^4 n\\
\partial_t \nabla^2 \phi + \{\phi,\nabla^2 \phi\} &= \alpha( \tilde \phi - \tilde n)  - \mu \nabla^2 \phi - D \nabla^6 \phi.
\end{align}
These equations use the usual normalizations $\ln(n/n_0) \to n, \phi \to e \phi /T_e, x \to \rho_s x, t \to t/\omega_{ci}.$ Here, $\{a, b\} \equiv \partial_y a \partial_x b - \partial_x a \partial_y b$ is the Poisson bracket, $\alpha$ is the ``adiabaticity parameter'' which measures the parallel electron response, $\mu=10^{-2}$  damps the flow at large scales, and the hyperdiffusion/hyperviscosity $D=10^{-4}$ removes energy at small scales. The small collisional terms are included primarily for numerical regularity --- turbulent transport dominates. We fix $\alpha=2$ throughout this work.

2-D HW is a representative paradigm for understanding the nonlinear dynamics of drift-wave turbulence. It is an appropriate testing ground for our method because (a) it captures the feedback between profile, flow, and turbulence field, allowing us to obtain a closed mean-field model, (b) training data can be generated easily, as simulations can be performed quickly on a high-performance machine, and (c) it is well-studied and relatively easy to treat analytically, allowing a means of checking our results.

We first perform direct numerical simulation of the 2-D HW system using the BOUT++ software \cite{bout} with a $512\times 512$ grid and Karniadakis' third-order splitting method \cite{karniadakis}. The box size corresponds to an effective $\rho_*$ of $1/51.5$. In the $x$-direction, we employ homogeneous Dirichlet boundary condition for $\phi$ and $\nabla^2 \phi$ and homogeneous Neumann boundary conditions for $n$. The system is periodic in the $y$-direction. 

To span a broad range of parameter space, we run 32 simulations, each with different initial conditions.  Ten simulations are initialized with a uniform background gradient ranging from $0.75 \le N' \le 3$ (this sets the system above the nonturbulent Dimits shift regime but below the strong turbulence regime). Seven simulations are initialized with a nonuniform background gradient $N'=\beta x/L_x$ with $1 \le \beta \le 5$. The remaining 15 simulations have both an initial uniform gradient $1\le N' \le 3$ and an initial background flow $V_y = v_0 \cos (2 \pi n x/L_x)$ with $n=1,2,3$. In all simulations, a small broad-spectrum fluctuation is initialized in the vorticity to start up the instability.

From the numerical solutions, we extract mean-field variables of interest and the corresponding particle flux $\Gamma = \langle  \tilde n \partial_y \tilde \phi \rangle$ and Reynolds stress $\Pi =  \langle \partial_x \tilde \phi \partial_y \tilde \phi \rangle$ at points in space and time. Data are outputted from simulation every $\Delta t =1$, from $t=10$ up to $t=2000$. The $x$-direction is coarse-grained into blocks of four points, over which any necessary finite differences are computed; thus, each simulation produces $128 \times 1990 = \num[group-separator = {,}]{254720}$ training data points. 

Finally, these data, and their images under reflection symmetry transformations, are used to train a DNN which outputs the flux as a function of chosen variables, using the Keras API \cite{keras} on top of TensorFlow \cite{tensorflow}. The DNN has the structure of a multilayer perceptron (MLP) with three hidden layers. Each hidden layer has eight units. As a reminder to the reader, an MLP is a simple network whose hidden layers successively transform the input like $\mathbf x^j \mapsto \sigma(\mathbf w^j_i \cdot \mathbf x^j + b^j_i),$ where $\sigma$ is a prescribed activation function, $\mathbf w_i^j$ is a trainable weight vector, $b_i^j$ is a trainable bias, $i$ refers to the index of the neuron, and $j$ refers to the index of the layer. This yields a total of 201 trainable parameters in our case. [See Ref.~\cite{ml_book} for an introduction, aimed at physicists, to DNNs and other machine learning methods.] The ``exponential linear unit'' \cite{elu}
\begin{equation}
f(x) = \begin{cases}
x, & x\ge0 \\
e^x-1, & x <0,
\end{cases}
\end{equation} 
 is used as the activation function for the hidden layers. No activation function is used for the output layer, since this is a regression (rather than classification) problem and there is no need to restrict the codomain.
 
Using standard methods, we train against the the loss function
\begin{equation}
L=\sum_{i}  \ln \left(\cosh (y_i^* - f_W(\mathbf{x}_i))\right) + \lambda ||W||^2,
\end{equation}
where $W$ is the matrix of network weights, $\mathbf{x}_i$ is the set of inputs ($U, N'$, etc.) for the $i$-th data point, $y_i^*$ is the corresponding flux, $f_W$ is the map encoded by the DNN which predicts the flux, $||\cdot||$ is the Frobenius norm, and $\lambda=10^{-5}$. We found this ``logcosh'' loss useful to suppress the effect of noise, as it is asymptotically \emph{linear} in the error for large arguments, but \emph{quadratic} (and smooth) for small arguments. The term $\lambda ||W||^2$ is the usual $L^2$ regularization, which reduces overfitting. Batch normalization \cite{ioffe} is applied after each hidden layer to accelerate training. 

The training procedure was repeated on ten random partitions of the data into training (80\% of the data) and validation sets (20\%). The results we quote are in fact the average of the outputs of the resulting ensemble of ten trained models. In separate training runs, we have checked that the model performs well on a test set that has been excluded from the training and validation sets. This test set corresponded to a specific range of initial $N'$ and comprised about 15\% of the data.

The 2-D HW model has a number of exact symmetries, which were found to be useful for training as a model constraint. The system is invariant under uniform shifts in both $n$ and $\phi$, as well as Galilean boosts in the poloidal direction. These continuous symmetries preclude explicit dependence of the fluxes on $\langle n \rangle$, $\langle \phi \rangle$, or the mean flow speed $V_y = - \partial_x \langle \phi \rangle$ in a local mean-field description. Moreover, we have a group of reflection symmetries with nontrivial elements
\begin{align}
&x \to -x, y\to -y; \\
&x \to -x, \phi \to -\phi, n \to -n ;\\
&y \to -y, \phi \to -\phi, n \to -n .
\end{align}
We approximately enforce these symmetries by duplicating and transforming the training data accordingly; for example, the first symmetry sends $\partial_x \langle n \rangle \to - \partial_x \langle n \rangle, \Gamma \to -\Gamma, \Pi \to \Pi$, etc. It may be possible to encode the symmetries in the structure of the DNN, but this is beyond the scope of the present work.

With the aid of the symmetry constraints, we train on the following set of independent variables: the mean density gradient $N'=\partial_x \langle n \rangle$, the mean vorticity $U=-\partial_x^2 \langle \phi \rangle$, $U'$, $U''$, and the turbulent potential enstrophy (PE) $\varepsilon = \langle (\tilde n -\nabla^2 \tilde \phi)^2 \rangle$ \cite{arash_pop}. The latter is a proxy for the turbulence intensity. While other choices are possible, the total PE $\varepsilon + (N+U)^2$ is conserved, so that the turbulent PE has the advantage of a dynamical description that is easy to write down.

\begin{figure}[htp]
\includegraphics[width=\columnwidth]{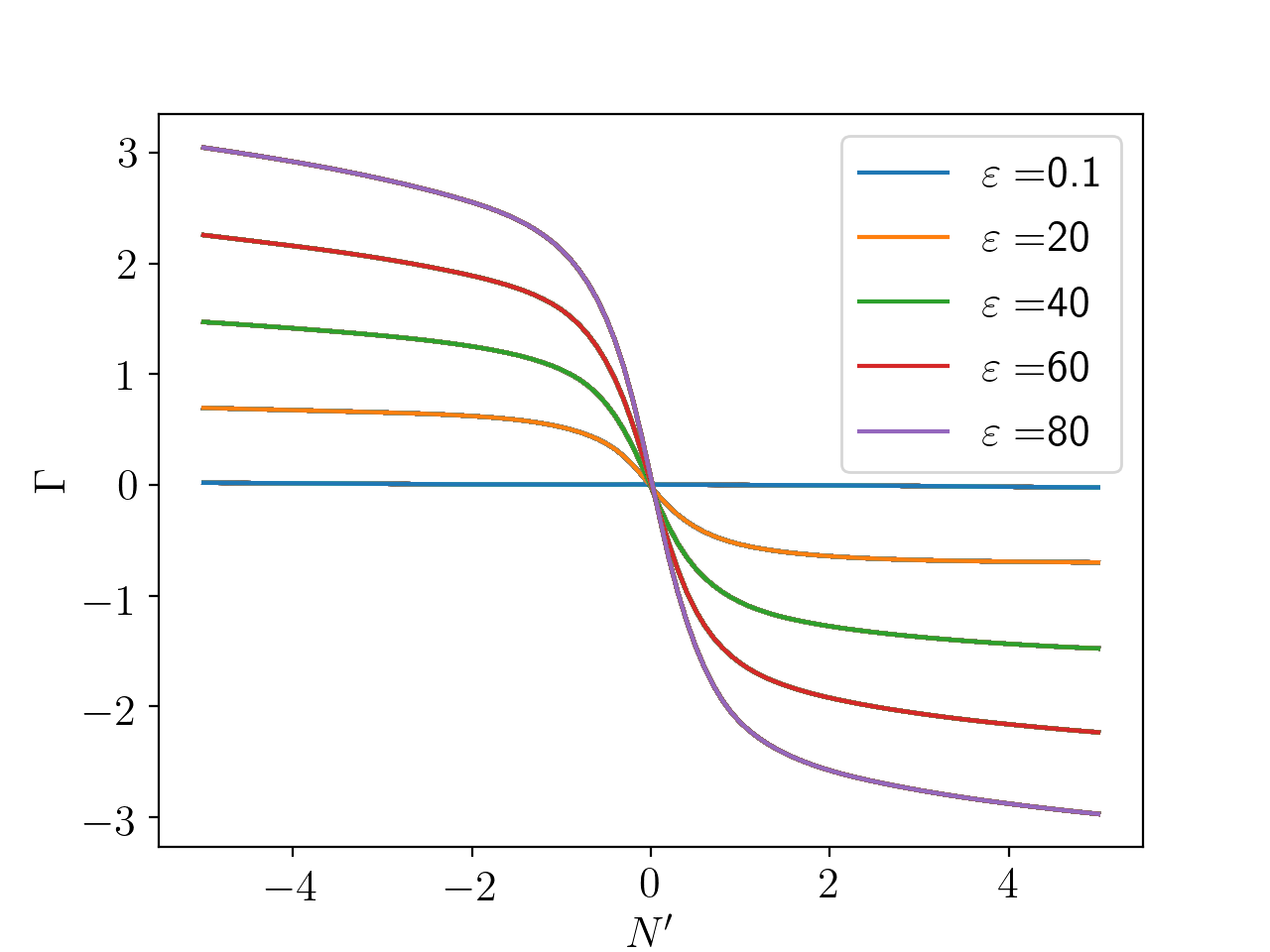}
\caption{Diffusive part of the learned particle flux, i.e. the flux at fixed $U=U'=U''=0$, as a function of $N'$ and $\varepsilon$. The dependence on $N'$ may be summarized as linear, plus saturation effects at large $N'$.}
\label{fig:diag}
\end{figure}

\begin{figure}[htp]
\includegraphics[width=\columnwidth]{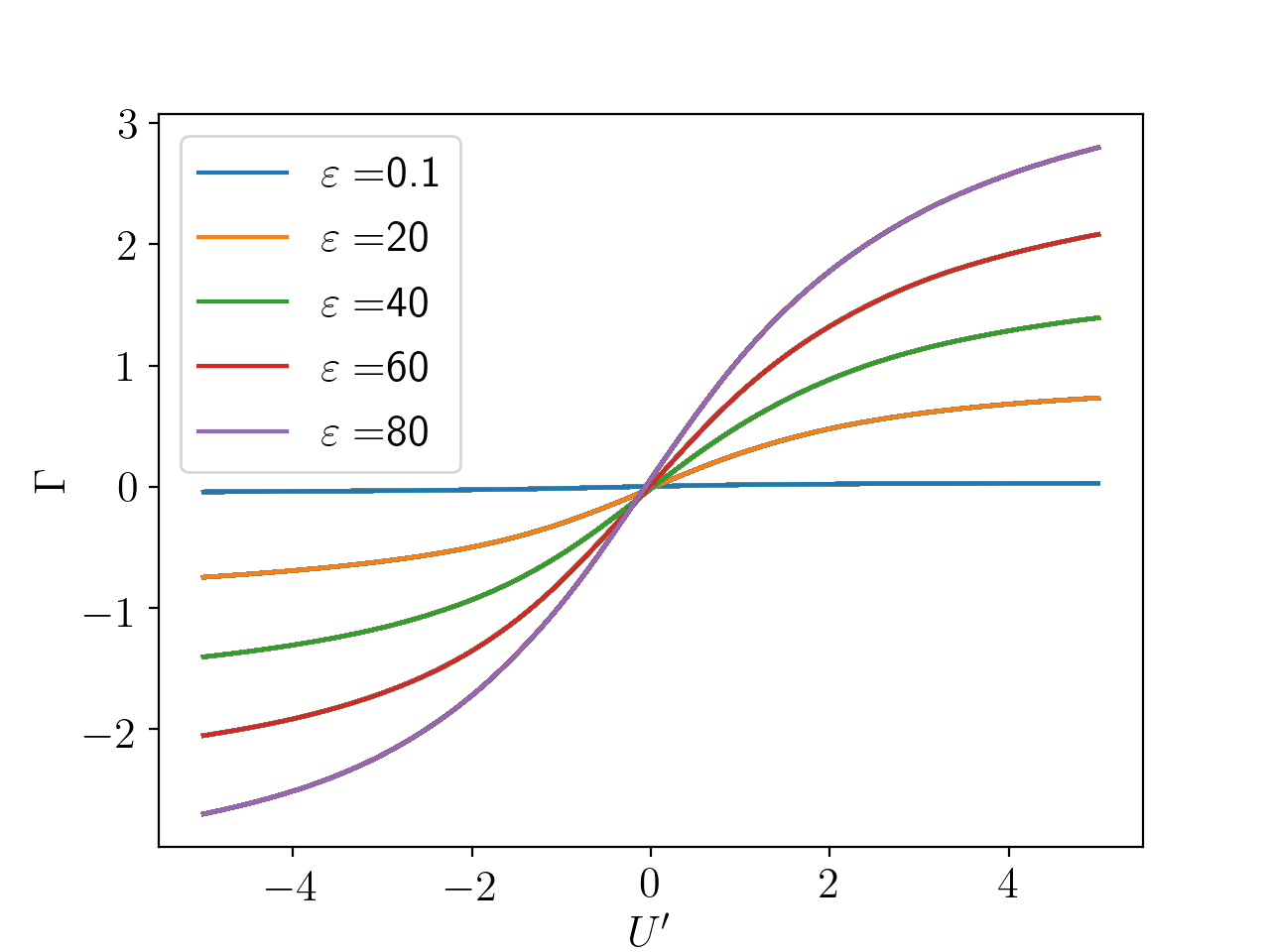}
\caption{Non-diffusive part of the learned particle flux,  i.e. the flux at fixed $N'=U=U''=0$, as a function of $U'$ and $\varepsilon$. Again, the dependence on $U'$ is roughly linear plus saturation effects.}
\label{fig:offdiag}
\end{figure}

\begin{figure}[htp]
\includegraphics[width=\columnwidth]{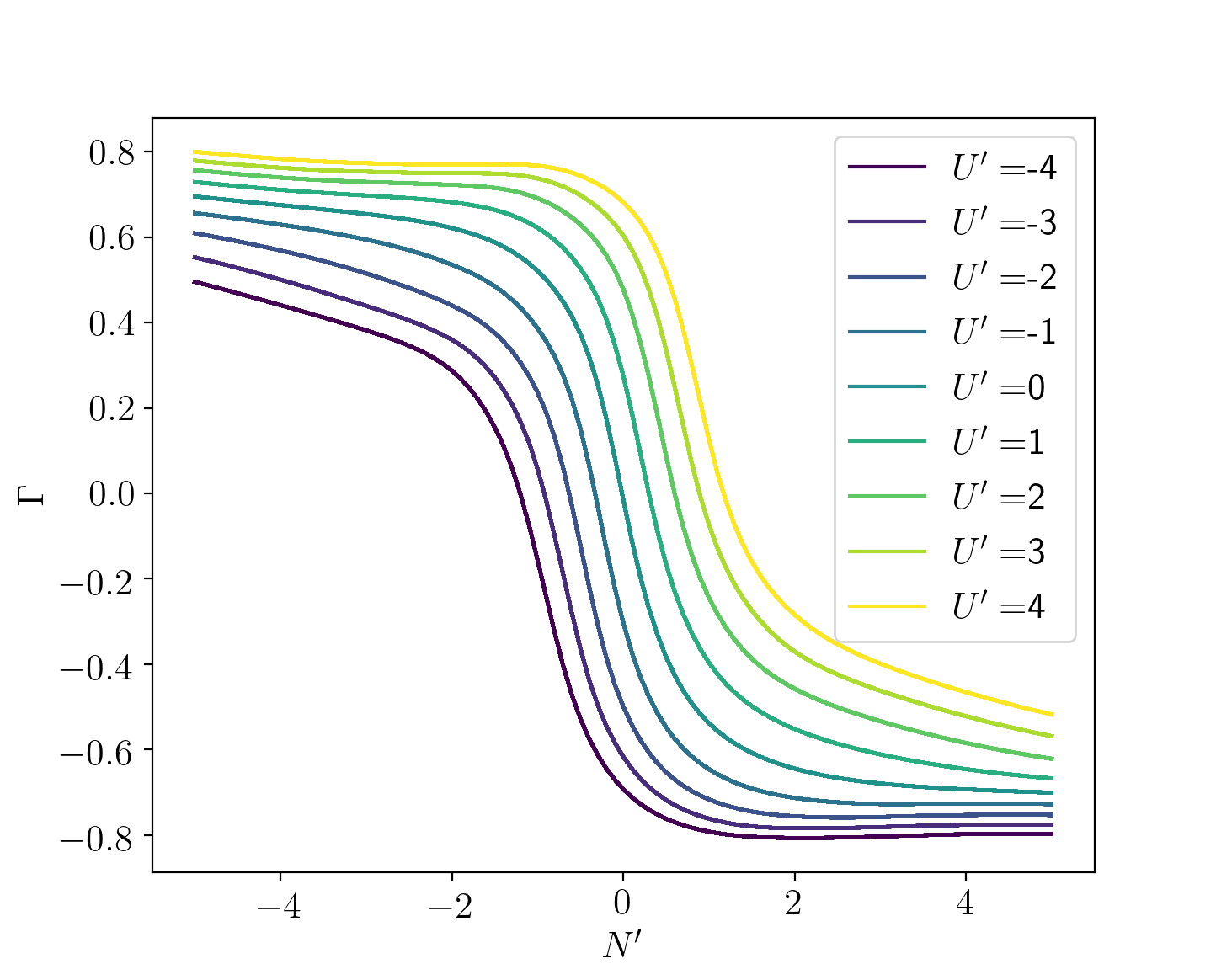}
\caption{Dependence of particle flux on both gradients: flux at fixed $U=U''=0$ and fixed $\varepsilon=20$, as a function of $N'$ and $U'$.}
\label{fig:both}
\end{figure}

The results for the particle flux are summarized in Figs.~\ref{fig:diag}--\ref{fig:both}. For $|N'|, |U'| \lesssim 1$, the flux is a linear combination of diffusive and nondiffusive terms.
\be
\Gamma \simeq \varepsilon(-D_n N' +D_u U'),
\ee
with $D_n\sim0.04$ and $D_u\sim0.015$. The first, diffusive term is the turbulent diffusion, which tends to relax the driving gradient. The second term is non-diffusive and is previously unreported. There are also higher-order saturation effects present at large $N'$ and $U'$. Not shown is the direct effect of the vorticity/shear $U$, which tends to reduce the flux independent of the sign of $U$. However, for typical values of the vorticity, this is a weak effect in this system ($\lesssim 10$\%).


The non-diffusive term, proportional to the vorticity gradient, will tend to corrugate the density profile in the presence of a quasiperiodic zonal flow, forming a staircase structure \cite{dif2010,dif2015,arash2019}. It can be recovered by a simple calculation that retains the background flow $V_y$. Due to the nonlinear convection of vorticity, the background flow shifts the drift-wave frequency:
\begin{align}
\mathrm{Re} \, \omega &= \frac{k_y (N' + V_y'')}{1+k^2} + k_y V_y \\
\mathrm{Im} \, \omega &= \frac{k_y^2}{\alpha (1+k^2)^3}(N' + V_y'') (k^2 N' -V_y''),
\end{align}
for $\alpha >1$. The coherent part of the particle flux is then straightforwardly computed as
\be
\Gamma \simeq -\frac1\alpha \int d^2 \mathbf{k} \, \frac{ k_y^2}{(1+k^2)^3}\left(k^2 N' -U'\right) \varepsilon_\mathbf{k},
\ee
where we have used $\varepsilon_\mathbf{k} \simeq (1+ k^2)^2 | \tilde \phi_\mathbf{k}|^2$.

\begin{figure}
\includegraphics[width=\columnwidth]{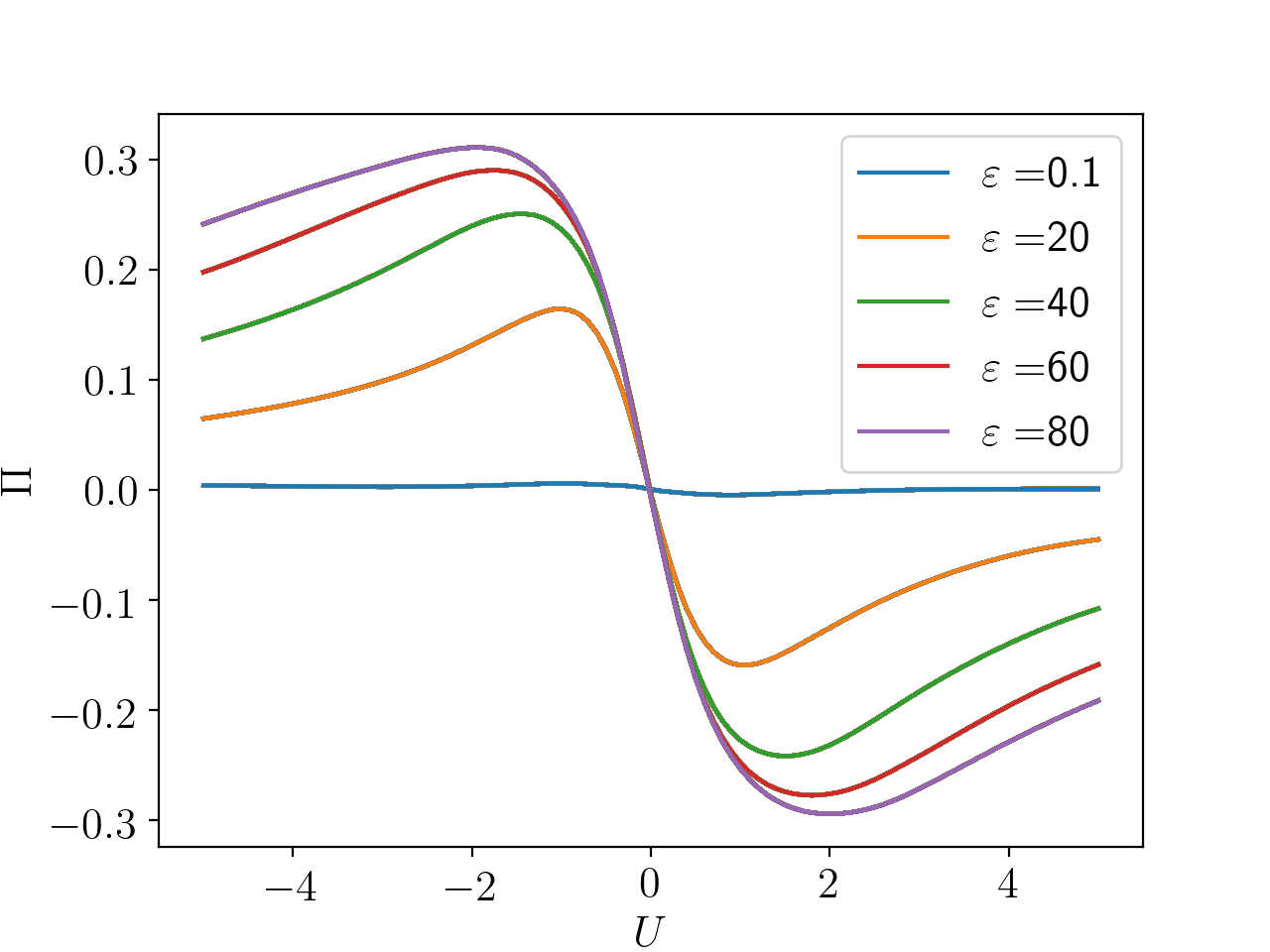}
\caption{Plot of learned Reynolds stress against vorticity $U,$ at fixed $N'=2$ and $U'=U''=0$ and several values of the intensity. Near $U=0,$ the behavior is that of a negative viscosity.}
\label{fig:reynolds}.
\end{figure}

\begin{figure}
\includegraphics[width=\columnwidth]{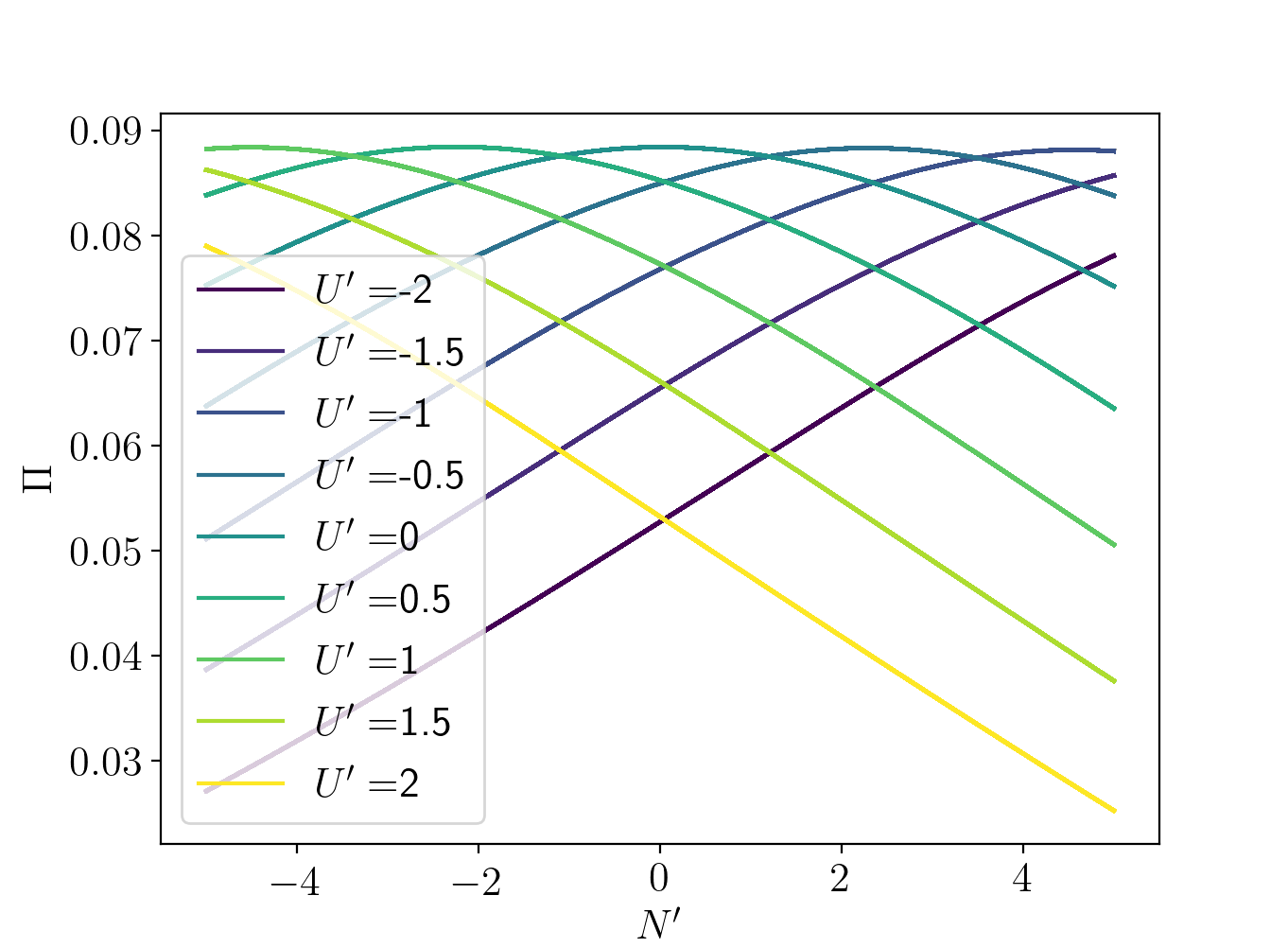}
\caption{Plot of learned Reynolds stress against $N'$ at fixed $U=1$,$\varepsilon=10$, $U''=0$, and several values of $U'$. The presence of a gradient in $U'$ or $N'$ tends to reduce the Reynolds stress.}
\label{fig:reynolds_both}.
\end{figure}

\begin{figure}
\includegraphics[width=\columnwidth]{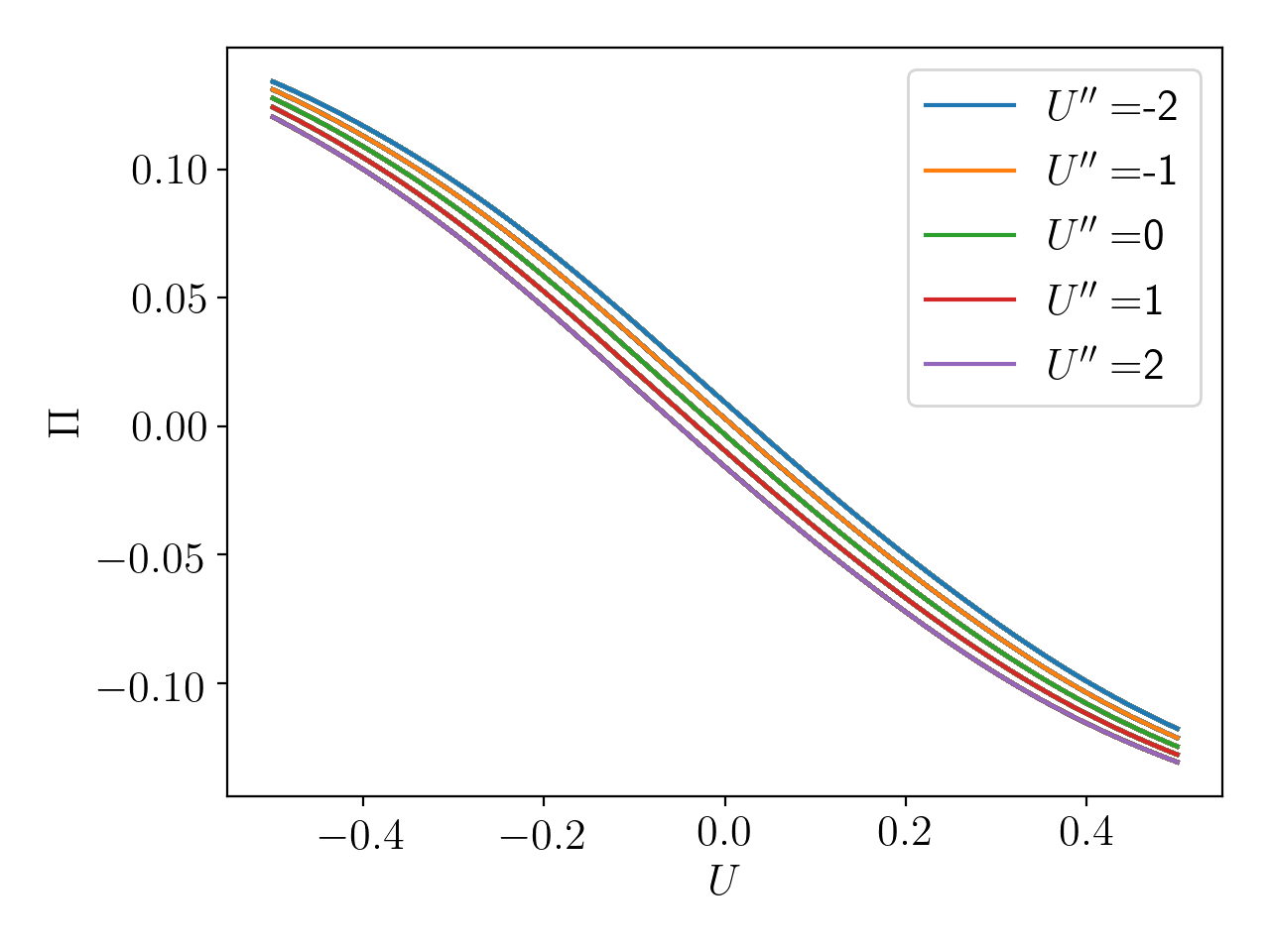}
\caption{Plot of learned Reynolds stress against vorticity $U$ at fixed $N'=2,$ $U'=0$, $\varepsilon=20$, and several values of $U''$. The leading order contribution from $U''$ is a stabilizing linear term.}
\label{fig:hyper1}.
\end{figure}

\begin{figure}
\includegraphics[width=\columnwidth]{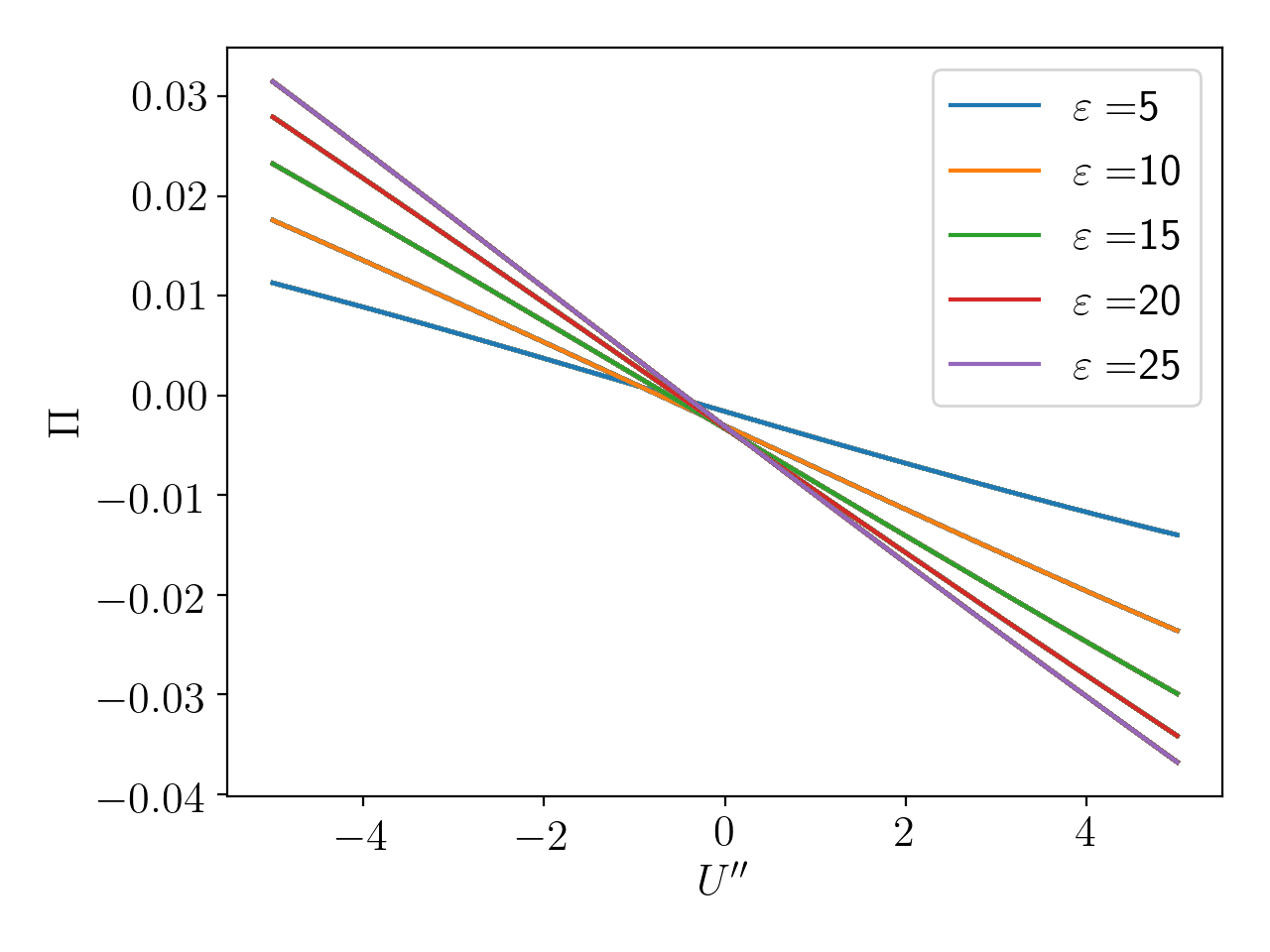}
\caption{Plot of learned Reynolds stress against $U''$ at fixed $N'=2,$ $U=U'=0$ and several values of the intensity. We should have $\Pi\to-\Pi$ under $U''\to-U''$ here, but the model fails to precisely learn this, which may be attributed to the relatively small contribution to the loss function from the hyperdiffusion term. However, it is clear that this term scales roughly as $\varepsilon U''$. }
\label{fig:hyper2}.
\end{figure}

Our results for the Reynolds stress indicate that zonal flows spontaneously generate by negative viscosity. For small $U$ and $\varepsilon$, the DNN obtains a model of the form
\be
\Pi \simeq \varepsilon f(N',U') (-\chi_1 U+\chi_3 U^3-\chi_4 \partial_x^2 U),
\ee
with $\chi_1\sim 0.015$ and $\chi_3\sim 0.01,$ $\chi_4 \sim 0.0005$, and
\be
f(N',U') \simeq \frac{1}{1+0.04(N' +4U')^2}.
\ee
These results are shown in Figs.~\ref{fig:reynolds}--\ref{fig:hyper2}.

Using the learned form of the Reynolds stress, the vorticity evolution $\partial_t U = \partial^2_x \Pi$ (neglecting dissipation) has the basic form of a Cahn-Hilliard equation \cite{cahn-hilliard} with dynamical coefficients. This agrees with previous theoretical work---see, for example, Ref.~\cite{diamond2005}. The negative viscosity $\chi_1$ destabilizes scales $\ell \gtrsim (\chi_4/\chi_1)^{1/2}$, and the cubic nonlinearity stabilizes large vorticities $U \gtrsim (\chi_1/\chi_3)^{1/2}.$ The stabilizing hyperviscous term is crucial for the stability of the vorticity evolution; in its absence, the zonal flow is unstable at all small scales and the dynamics are ill-posed. That the DNN recovers this small term shows that the method passes a sensitive test. 

The prefactor $f$ is a new, higher-order effect which further stabilizes the growth of the zonal flow; as $U'$ steepens due to the negative viscosity, the denominator of $f$ increases, which in turn reduces the Reynolds stress and inhibits further steepening. The DNN also finds higher-order, saturating terms in $U$ which result in power-law decay of the Reynolds stress with $U$. 

The equations 
\begin{align}
\partial_t N + \partial_x \Gamma &=0 \\
\partial_t U - \partial_x^2 \Pi &= 0, \\
\end{align}
equipped with the models for $\Gamma$ and $\Pi$ learned by the DNN, can be coupled with a model for the evolution for the turbulence intensity to obtain a reduced, three-field 1-D model for the turbulence dynamics. An appropriate model equation is 
\be
\partial_t \varepsilon + 2(\Gamma - \partial_x \Pi)(N'+U')\varepsilon = -\gamma_0 \varepsilon - \gamma_{NL} \varepsilon^2.
\ee
This equation expresses conservation of the potential enstrophy $W= \int d^2\mathbf{x} \, (n-\nabla^2 \phi)^2$, equivalent to the mean square charge density fluctuation. It can be derived either by integrating the wave-kinetic equation (WKE) over reciprocal space or by manipulating the equations of evolution of $\tilde n$ and $\nabla^2 \tilde \phi$ \cite{arash_pop} and neglecting the flux of turbulent PE, which models spreading. The linear damping $\gamma_0=D_n \kappa_0^2$  is necessary to model the threshold density gradient for linear instability $\kappa_0$. The nonlinear damping $\gamma_{NL}$ models the transfer to dissipation via the cascade. 

The closed system for $N, U$ and $\varepsilon$ captures the initial growth of turbulence, the spontaneous formation of a zonal flow, and the back reaction on the profile. We will solve it numerically and compare to DNS of the 2-D system in a forthcoming paper.

We have thus used the deep learning method to extract a simple mean-field model for the drift-wave/zonal flow system directly from numerical solution data. The only other inputs are exact symmetries and the choice of mean field parameters ($N'$, $U$, $\varepsilon$, etc.). The method successfully reproduces previous analytical results for the Reynolds stress, including the negative viscosity effect and crucial terms which regularize it. The analogy to the Cahn-Hilliard equation, which models spinodal decomposition of a mixture, has a clear physical interpretation: positively and negatively signed vortices spontaneously separate.

Moreover, the method recovers a new, non-diffusive particle flux driven by the gradient of vorticity, in addition to the well-known diffusive flux. The coupling to vorticity gradient is significant, of the same order of magnitude as the density gradient coupling, and far stronger than the direct coupling to the shear. The physical origins of the non-diffusive effect are in the nonlinear convection of vorticity, which shifts the drift wave frequency. It has clear implications for structure formation, as it tends to corrugate the density profile. The formation of staircase-like structures in the profile is well-known \cite{dif2010,dif2015,arash2019}, but the mechanism highlighted in this work is distinct from previous models based on bistability \cite{arash_pre,arash_pop,guo2019}.

On the other hand, our method has a number of limitations. The assumption of spatial and temporal locality is, while standard, \emph{ad hoc} and quite severe. In reality, the spectral structure of the turbulence, implicitly taken here to be constant in time, will carry some memory of the time history. Moreover, spatially nonlocal transport models have seen some success \cite{dif2010,ida}. The mean-field approximation, too, is only reasonable in the weak turbulence limit, wherein the flow retains its axisymmetry and intermittency effects are relatively insignificant.

While, in this work, the structure of the DNN model was simple enough to interpret graphically, in other, more complex applications, peering into the ``black box'' will likely pose a greater challenge, and more sophisticated methods may be necessary. Adjudicating the ``correctness'' of this structure is another challenge altogether that requires physics intuition. In this work, our confidence in the results rests primarily on (a) their respect for underlying symmetries, (b) their respect for the physical constraint that the fluxes must vanish at $\varepsilon \to 0,$ (c) their agreement with analytical calculations, and (d) their robustness to variations in the training data.

Collisional and neoclassical contributions to the fluxes were neglected in this work, with collisional terms set deliberately small. In a real system, these contributions may be significant and/or have complex structure.

We anticipate our deep learning approach may be straightforwardly applied to other turbulent systems with quasisymmetry along at least one spatial degree of freedom. Future work will focus on such applications, as well as relaxing the assumption of space-time locality. 

\acknowledgments{We acknowledge Arash Ashourvan, Norman Cao, Guilhem Dif-Pradalier, Ozg\"ur G\"urcan, and T. S.\ Hahm for useful discussions, many of which took place at the 2017 and 2019 Festivals de Th\'eorie in Aix-en-Provence and the 2018 Chengdu Theory Festival. This work used the Extreme Science and Engineering Discovery Environment (XSEDE) \cite{xsede}, which is supported by National Science Foundation grant number ACI-1548562, using the Comet cluster at the San Diego Supercomputing Center (SDSC) through allocation TG-PHY190014. It was supported by the U.S. Department of Energy, Office of Science, Office of Fusion Energy Sciences under Award Number DE-FG02-04ER54738.}

%

\end{document}